\documentstyle[psfig]{mn}

\newif\ifAMStwofonts

\def\simlt{\lower.5ex\hbox{$\; \buildrel < \over \sim \;$}}
\def\simgt{\lower.5ex\hbox{$\; \buildrel > \over \sim \;$}}

\ifoldfss
  \ifCUPmtlplainloaded \else
    \NewTextAlphabet{textbfit} {cmbxti10} {}
    \NewTextAlphabet{textbfss} {cmssbx10} {}
    \NewMathAlphabet{mathbfit} {cmbxti10} {} 
    \NewMathAlphabet{mathbfss} {cmssbx10} {} 
  \fi
  \ifAMStwofonts
    \ifCUPmtlplainloaded \else
      \NewSymbolFont{upmath} {eurm10}
      \NewSymbolFont{AMSa} {msam10}
      \NewMathSymbol{\upi}     {0}{upmath}{19}
      \NewMathSymbol{\umu}     {0}{upmath}{16}
      \NewMathSymbol{\upartial}{0}{upmath}{40}
      \NewMathSymbol{\leqslant}{3}{AMSa}{36}
      \NewMathSymbol{\geqslant}{3}{AMSa}{3E}

    \fi
  \fi
\fi 

\ifnfssone
  \newmathalphabet{\mathit}
  \addtoversion{normal}{\mathit}{cmr}{m}{it}
  \addtoversion{bold}{\mathit}{cmr}{bx}{it}
  \newmathalphabet{\mathbfit} 
  \addtoversion{normal}{\mathbfit}{cmr}{bx}{it}
  \addtoversion{bold}{\mathbfit}{cmr}{bx}{it}
  \newmathalphabet{\mathbfss} 
  \addtoversion{normal}{\mathbfss}{cmss}{bx}{n}
  \addtoversion{bold}{\mathbfss}{cmss}{bx}{n}
  \ifAMStwofonts
    \ifCUPmtlplainloaded \else
      \UseAMStwoboldmath
      \makeatletter
      \new@mathgroup\upmath@group
      \define@mathgroup\mv@normal\upmath@group{eur}{m}{n}
      \define@mathgroup\mv@bold\upmath@group{eur}{b}{n}
      \edef\UPM{\hexnumber\upmath@group}
      \new@mathgroup\amsa@group
      \define@mathgroup\mv@normal\amsa@group{msa}{m}{n}
      \define@mathgroup\mv@bold\amsa@group{msa}{m}{n}
      \edef\AMSa{\hexnumber\amsa@group}
      \makeatother
      \mathchardef\upi="0\UPM19
      \mathchardef\umu="0\UPM16
      \mathchardef\upartial="0\UPM40
      \mathchardef\leqslant="3\AMSa36
      \mathchardef\geqslant="3\AMSa3E
    \fi
  \fi
\fi 

\ifnfsstwo
  \DeclareMathAlphabet{\mathbfit}{OT1}{cmr}{bx}{it}
  \SetMathAlphabet\mathbfit{bold}{OT1}{cmr}{bx}{it}
  \DeclareMathAlphabet{\mathbfss}{OT1}{cmss}{bx}{n}
  \SetMathAlphabet\mathbfss{bold}{OT1}{cmss}{bx}{n}
  \ifAMStwofonts
    \ifCUPmtlplainloaded \else
      \DeclareSymbolFont{UPM}{U}{eur}{m}{n}
      \SetSymbolFont{UPM}{bold}{U}{eur}{b}{n}
      \DeclareSymbolFont{AMSa}{U}{msa}{m}{n}
      \DeclareMathSymbol{\upi}{0}{UPM}{"19}
      \DeclareMathSymbol{\umu}{0}{UPM}{"16}
      \DeclareMathSymbol{\upartial}{0}{UPM}{"40}
      \DeclareMathSymbol{\leqslant}{3}{AMSa}{"36}
      \DeclareMathSymbol{\geqslant}{3}{AMSa}{"3E}
    \fi
  \fi
\fi 

\ifCUPmtlplainloaded \else
  \ifAMStwofonts \else 
    \def\upi{\pi}
    \def\umu{\mu}
    \def\upartial{\partial}
  \fi
\fi

\title[Binary parameters of V2051~Ophiuchi]
		{HST and ground-based eclipse observations of V2051~Ophiuchi:
		Binary parameters\thanks{Based on observations with the NASA/ESA
		{\em Hubble Space Telescope}, obtained at the Space Telescope Science
		Institute, which is operated by the Association of Universities for
		Research in Astronomy, Inc., under NASA contract NAS~5-2655, and on
		observations made at Laborat\'{o}rio Nacional de Astrof\'\i sica/CNPq,
		Brazil.}.}
\author[R. Baptista et~al.]
       {Raymundo Baptista$^1$, M.\,S. Catal\'an$^2$, Keith Horne$^3$ and D. Zilli$^1$ \\
       $^1$ Departamento de F\'\i sica, Universidade Federal de Santa Catarina,
       Campus Trindade, 88040-900, Florian\'opolis - SC, Brazil, \\
       ~ email: bap@fsc.ufsc.br, zilli@fsc.ufsc.br \\
       $^2$ Department of Physics, Keele University, Keele, Staffordshire,
       ST5 5BG, UK, email: msc@astro.keele.ac.uk \\
       $^3$ School of Physics \& Astronomy, University of St.\,Andrews,
       North Haugh, St.\,Andrews, Fife, KY16 9SS, Scotland, \\
       ~ email: kdh1@st-and.ac.uk }
\date{Accepted 1998 June 8; Received 1998 January 5}

\pagerange{1--16}
\pubyear{1998}

\begin{document}

\label{firstpage}

\maketitle

\begin{abstract}

We report on high-speed eclipse photometry of the dwarf nova V2051~Oph
while it was in a low brightness state, at $B \simeq 16.2$ mag. In
comparison to the average IUE spectra, the ultraviolet continuum and
emission lines appear reduced by factors of, respectively, $\simeq 4$
and $\simeq 5$.  Flickering activity is mostly suppressed and the
lightcurve shows the eclipse of a compact white dwarf at disc centre
which contributes $\simeq 60$ per cent of the total light at 3900--4300
\AA.

We use measurements of contact phases in the eclipse lightcurve to
derive the binary geometry and to estimate masses and relevant
dimensions. We find a mass ratio of $q= 0.19 \pm 0.03$ and an
inclination of $i= 83 \pm 2$ degrees.  The masses of the component
stars are $M_1 = 0.78\pm 0.06 \; M_\odot$ and $M_2 = 0.15 \pm 0.03 \;
M_\odot$.  Our photometric model predicts $K_1 = 83\pm 12$ km s$^{-1}$
and $K_2= 435 \pm 11$ km s$^{-1}$.  The predicted value of K$_1$ is in
accordance with the velocity amplitude obtained from the emission lines
after a correction for asymmetric line emission in the disc is made
(Watts et~al. 1986).

The secondary of V2051~Oph is significantly more massive than the
secondaries of the other ultra-short period dwarf novae.  V2051~Oph is
probably a relatively young system, whose secondary star had not enough
time to evolve out of thermal equilibrium.

\end{abstract}


\section{Introduction}

Cataclysmic Variables (CVs) are close binary systems in which
mass transfer occurs from a late-type star filling its Roche lobe
to a companion white dwarf via an accretion disc or column (Warner
1996). The subclass of the dwarf novae comprises systems in which the
white dwarf is non-magnetic (B$\simlt 10^4$~G) and which are mostly
known for their recurrent, large amplitude outbursts ($\Delta m =
3-5$~mag, typical duration of 5-10 days).  These outbursts reflects a
change in the structure of their discs -- from a cool, optically thin,
low viscosity state to a hot, optically thick, high viscosity state
-- and which are usually parameterized as a large change in the mass
accretion rate ( \.{M}$= 10^{-11} \; M_\odot \; yr^{-1} \mapsto
10^{-9} \; M_\odot\; yr^{-1}$.  See, e.g.  Pringle, Verbunt \&
Wade 1986).  In quiescence, when the accretion disc is faint, the white
dwarf and the bright spot (where the stream of transferred matter hits the
edge of the accretion disc) dominate the optical and ultraviolet light
from the system, and can be seen as sharp changes in the slope of the
lightcurve in eclipsing systems.  From the timings of these features,
a precise determination of the binary parameters can be obtained.

V2051~Oph is an eclipsing dwarf nova with an orbital period of 90
minutes. High speed photometry by Warner (1983) and Cook \& Brunt (1983)
shows that the orbital lightcurve is characterized by deep eclipses
($\Delta B\simeq 2.5$ mag) and large amplitude flickering ($\simgt
30$ per cent), responsible for a variety of eclipse morphologies.
Significant ($> 50$ per cent) brightness changes occur on timescales
shorter than the orbital period but no stable orbital hump (associated
to anisotropic emission from a bright spot) precedes the eclipse. Unlike
most dwarf novae, which usually display an orbital hump with maximum at
phase $\sim 0.8$, in V2051~Oph the humps apparently occur at any orbital
phase (Berriman, Kenyon \& Bailey 1986).  The compact source usually seen
at disc centre is variable both in brightness and in the duration of its
immergence and emergence (from $\simeq 50$ to 80 s), and was interpreted
by Warner \& Cropper (1983) to be the inner region of the disc and not
the white dwarf.  Only one relatively short outburst of V2051~Oph has
been reported since its discovery, in which it reached $B\simeq 13$ mag
(Warner \& O'Donoghue 1987).
The analysis of optical and infrared lightcurves led Berriman et~al.
(1986) to suggest that the accretion disc in V2051~Oph consists
largely of optically thin material, in accordance with the observed color
changes during eclipse (Cook \& Brunt 1983).  V2051~Oph was proposed to be
a low field polar system by Warner \& O'Donoghue (1987) -- a possibility
first mentioned by Bond \& Wagner (1977) -- based on the interpretation
of their eclipse maps and on the observation of a 42\,s oscillation in
the optical during outburst, reminiscent of the rapid oscillations seen
in polars.  However, in quiescence the emission lines are double peaked
and exhibit the classical rotational disturbance effect during eclipse,
suggesting the existence of a prograde accretion disc (Cook \& Brunt 1983;
Watts et~al. 1986).
Watts et~al. (1986) combined the measured $K_1$ value with
velocity-dependent eclipse data of the emission lines to derive a
dynamical solution for the binary, obtaining $M_1= 0.43 \; M_\odot$,
$M_2= 0.11 \; M_\odot$ and a radius $R_2= 0.15 \; R_\odot$. They also fit
the ultraviolet to optical spectrum with a steady-state optically thick
disc model to find \.{M}$= 1.3 \times 10^{-10}\; M_\odot \; yr^{-1}$
and a distance to the system between 90 and 150 pc.

V2051~Oph belongs to a small group of ultra-short period eclipsing dwarf
nova together with Z~Cha, OY~Car and HT~Cas. Binary parameters for those
other systems have been determined with high precision using the method
described by Cook \& Warner (1984) to derive the mass ratio, inclination
and white dwarf radius from the white dwarf and bright spot eclipse phases.
In V2051~Oph, the absence of a well defined bright spot and the uncertainty
in associating the compact central source with the white dwarf (due to its
variability both in width and brightness) has, until now, precluded the
use of this method to derive its binary parameters.

In this paper we present and discuss Hubble Space Telescope and
ground-based eclipse observations while V2051~Oph was in an occasional
low brightness state. The data are described in section\,\ref{data}.
In section\,\ref{param} we use contact phases of the white dwarf
and bright spot to develop a purely photometric model for the binary,
deriving masses and radii of both stars, the binary inclination and the
orbital separation. Section\,\ref{discussion} discusses possible systematic
effects affecting the derived binary parameters, possible causes for the
observed low brightness, the secondary star and the evolutionary stage
of V2051~Oph.  Our results are summarized in section\,\ref{fim}.

\section{Observations} \label{data}

\subsection{HST/FOS high-speed spectroscopy}

The {\em Faint Object Spectrograph} (FOS) onboard the {\em Hubble Space
Telescope} (HST) was used to obtain time-resolved spectroscopy of two
consecutive eclipses of the dwarf nova V2051~Oph on 1996 January 29.
The observations, summarized in Table\,\ref{log}, consist of two series
of exposures in the RAPID readout mode at a time resolution of 3.38\,s
and a dead-time between exposures of 0.1\,s.  During data acquisition,
the spectrum was shifted electronically by 1/2 of a diode along the 516
diode Digicon array (`1/2 sub-stepping') resulting in a 1032-pixel
spectrum with a net exposure time of 1.64\,s pixel$^{-1}$.  The first run
(H1) was carried out using a $0.86''$ circular aperture and the G400H
grating (spectral resolution $\Delta\lambda= 1.5$ \AA\ pixel$^{-1}$)
and covered the egress of the first eclipse.  The second run (H2)
was performed using a $0.86''\times 0.86''$ square aperture and the
G160L grating ($\Delta\lambda= 3.5$ \AA\ pixel$^{-1}$) and covered the
following eclipse.
The spectral range covered by the HST observations is, respectively for
the G160L and the G400H data, $1150-2500$ \AA\ and $3200-4800$ \AA.

The observations were reduced using procedures similar to the standard
STSDAS pipeline, and included flat-field and geomagnetically induced
motion (`GIMP') corrections, background and scattered light subtraction,
wavelength and absolute flux calibrations.  The absolute photometric
accuracy of these observations should be about 4 per cent.

For the G160L observations, the undispersed order-zero light is also
recorded on the diode array, providing a broad-band UV-optical photometric
measurement that is simultaneous with each of the ultraviolet spectra.
The order-zero light has a passband with a FWHM of 1900 \AA\ and a pivot
wavelength of 3400 \AA. To convert counts to flux densities we used
the predicted post-COSTAR response of 950 counts s$^{-1}$ mJy$^{-1}$
of Eracleous \& Horne (1994), which has un uncertainty of 50 per cent.

\begin{table*}
 \centering
 \begin{minipage}{120mm}
  \caption{Journal of the observations.}
  \label{log}
\begin{tabular}{@{}lccccccc@{}}
~~date & run & Start & $\Delta t$ & No. of & Spectral & Phase range &
Instrument \\ [-0.5ex]
(1996) &  & (UT) & (s) & exposures & range & (cycles) \\ [1ex]
Jan 29 & H1 & 19:03 & 3.38 & 113 & 3226-4781 \AA & $+0.01, +0.08$ & HST/FOS/BL \\
Jan 29 & H2 & 20:24 & 3.38 & 693 & 1150-2507 \AA & $-0.09, +0.34$ & HST/FOS/BL \\
[1ex]
Jun 15 & L1 & 22:21 & 5.0 & 547 &  B-band     & $-0.24, +0.27$ & FOTRAP \\
Jun 16 & L2 & 22:24 & 5.0 & 398 & white light & $-0.18, +0.18$ & FOTRAP \\
Jun 16 & L3 & 23:45 & 5.0 & 516 & white light & $-0.29, +0.19$ & FOTRAP \\
Jun 17 & L4 & 22:19 & 5.0 & 432 & white light & $-0.23, +0.17$ & FOTRAP \\
\end{tabular}
\end{minipage}
\end{table*}

\begin{figure*}
\centerline{\psfig{figure=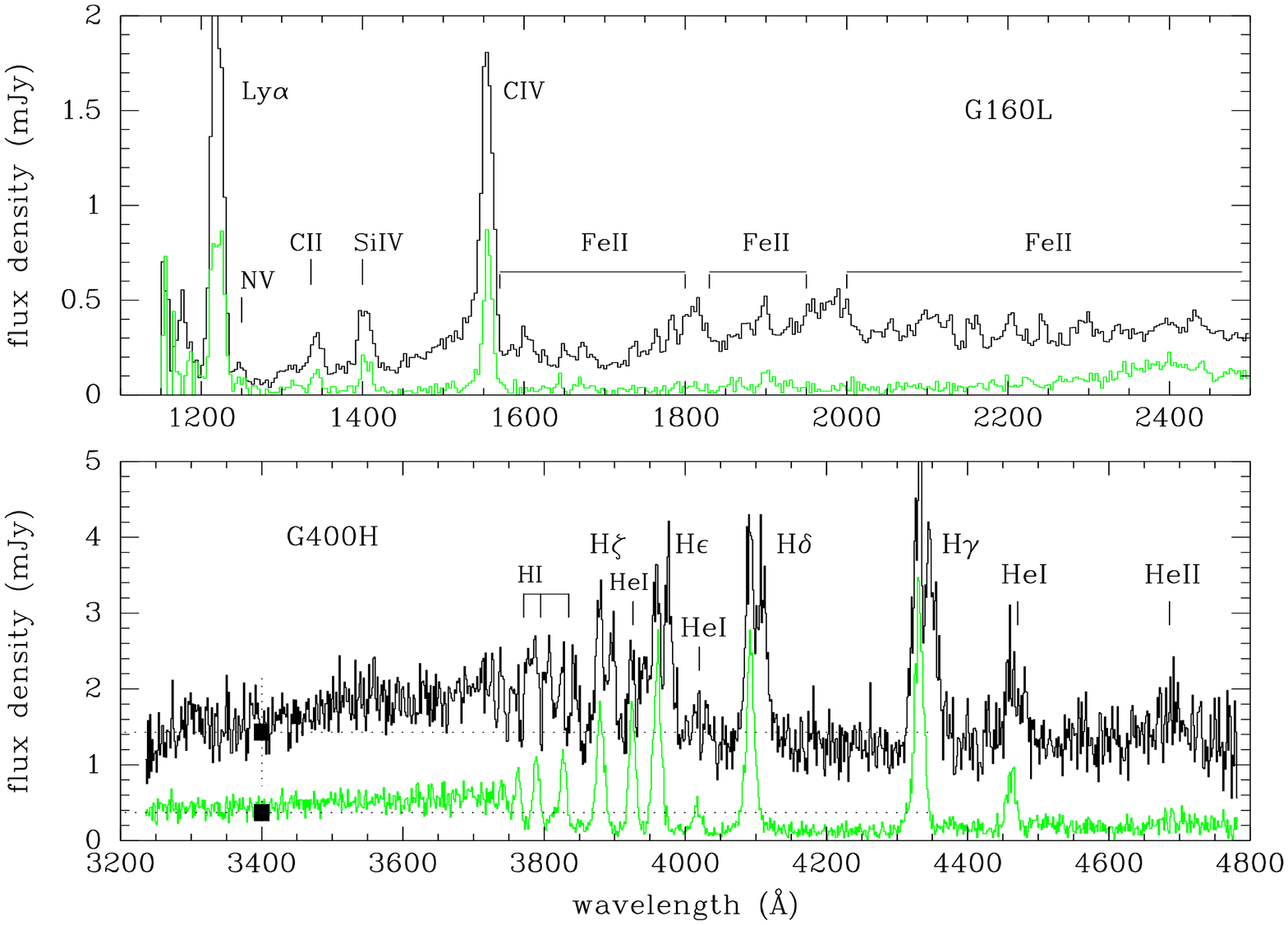,angle=-90,width=18cm,rheight=15cm}}
\caption{Average out-of-eclipse (black) and mid-eclipse (light gray) HST
spectra of V2051 Oph. Major emission and absorption features are labelled.
Phase ranges of each spectrum are as follows: +0.1 to +0.3 (G160L, out of
eclipse), $-0.02$ to +0.03 (G160L, mid-eclipse), +0.082 to +0.09 (G400H,
out-of-eclipse), and +0.02 to +0.03 (G400H, mid-eclipse).
Filled squares in the G400H panel indicate the fluxes of the G160L order
zero light computed at the same phase ranges as the G400H average spectra.
Horizontal dotted lines show the FWHM of the G160L order zero passband. }
\label{fig1}
\end{figure*}

Average out-of-eclipse and mid-eclipse spectra of the G160L data are
shown in the upper panel of Fig.\,\ref{fig1}. The most prominent lines
and bands are labeled.
The comparison of the G160L out-of-eclipse spectrum with the average IUE
spectrum of Watts et~al. (1986) reveals that V2051~Oph was considerably
fainter than usual at the time of the HST observations. The flux in the
continuum and the emission lines have been reduced by factors of,
respectively, $\simeq 4$ and $\simeq 5$.
The lower panel of Fig\,\ref{fig1} shows average out-of-eclipse and
mid-eclipse spectra of the G400H data.
The G160L order-zero passband overlaps the spectral coverage of the G400H
data.
Filled squares show the fluxes of the G160L order zero light computed
at the same orbital phase ranges as the G400H average spectra.
This comparison indicates that V2051~Oph remained at roughly the same
brightness level during the HST observations.
%

Lightcurves at selected passbands for both HST runs are shown in
Fig.\,\ref{fig2}. These will be discussed in detail in section\,\ref{cluz}.

\subsection{high-speed photometry}

High-speed photometry of V2051~Oph was obtained with the 1.6\,m
telescope of Laborat\'orio Nacional de Astrof\'\i sica (LNA/CNPq),
in southern Brazil, on 1996 June 15-17.
Four eclipses were collected, one in the B-band and the remaining
in white light (W). These observations are detailed in Table\,\ref{log}.

The observations were performed with the one-channel FOTRAP photometer
in its single-filter mode at a time resolution of 5\,s, using a 11.3''
diaphragm. A Hamamatsu R943-02 PMT was employed.  Sky measurements were
taken at intervals of 15-20 min at a position 20'' to the north of the
variable, except during eclipse.  The sky measurements were fitted by
a low-order polynomial and then subtracted from the raw data.  A close
comparison star was also regularly observed to check for the presence
of clouds and sky transparency variations. From these observations,
we confirmed that all 4 runs were performed under good sky conditions.
Coefficients of extinction and of transformation to the UBVRI standard
system were derived from observations of E-regions stars of Graham
(1982) and blue spectrophotometric standards of Stone \& Baldwin (1983).
The reader is referred to Jablonski et al. (1994) for a detailed study
of the reliability of the transformations from FOTRAP's natural to the
standard UBVRI system, in particular for objects of peculiar spectra
such as CVs. We used the relation of Lamla (1981), $B= 2.5 \log f_B {\rm
(mJy)} - 16.57$, to transform $B$ magnitudes to flux density units. The
absolute calibration of the B-band observations is accurate to better
than 4 per cent.

The ground-based lightcurves of V2051~Oph are shown in Fig.\,\ref{fig3}
and will be discussed in section\,\ref{cluz}.

\subsection{New eclipse timings}

Mid-eclipse times of V2051~Oph were measured by employing the bissected
chord method and by computing the mid-point of the phases of maximum
and minimum derivative in the lightcurves (section\,\ref{param}). These
methods yield values which are consistent with each other under the
uncertainties, and therefore we adopted the mean as the mid-eclipse time.
The new heliocentric (HJD) as well as baricentric (BJED) timings of
V2051~Oph are listed in Table\,\ref{timings} with corresponding cycle
number and uncertainties (quoted in parenthesis).  Also shown are the
(O--C) values with respect to the revised linear ephemeris of Baptista,
Tripplet \& Bond (1998),
\begin{equation}
T_{mid} = BJED \;\; 2\,443\,245.977\,850 + 0.062\,427\,8595 \times E
\label{efem}
\end{equation}
The measured (O--C) values are in accordance with the cyclical variations
in the orbital period of V2051~Oph observed by Baptista, Tripplet \&
Bond (1998).
%
%
\begin{table}
\begin{minipage}{80mm}
  \caption{New eclipse timings.} \label{timings}
\begin{tabular}{@{}cccc@{}}
Cycle  & HJD & BJED & ~(O$-$C) \footnote{Observed minus calculated times
with respect to the ephemeris of eq.\ref{efem}.} \\ [-0.5ex]
     & (2,450,000+) & (2,450,000+) & (cycles) \\ [1ex]
109\,988 & 112.29277 & 112.29273(4) & $-0.0084$ \\ 
109\,989 & 112.35517 & 112.35513(4) & $-0.0089$ \\
112\,201 & 250.44558 & 250.44555(2) & $-0.0090$ \\
112\,217 & 251.44443 & 251.44440(1) & $-0.0089$ \\
112\,218 & 251.50688 & 251.50685(1) & $-0.0086$ \\
112\,233 & 252.44331 & 252.44328(2) & $-0.0084$ \\ [-4ex]
\end{tabular}
\end{minipage}
\end{table}

\subsection{Eclipse lightcurves} \label{cluz}

We adopted the following convention regarding the phases: conjunction
occurs at phase zero, the phases are negative before conjunction and
positive afterwards.  The lightcurves were phased according to the
linear ephemeris of eq.\,\ref{efem}.  A phase correction of +0.0087
cycles was further applied to the data to make the centre of the white
dwarf eclipse coincident with phase zero (see section\,\ref{param}).

V2051~Oph was particularly faint at the time of both the HST and the
ground-based observations. From the B-band lightcurve, we estimate a
magnitude of $B= 16.3$ mag, while the G400H data yield $B \simeq$ 16.2
mag. This is considerably fainter than the $B= 14.7-15.5$ mag range
previously reported in the literature (Warner \& Cropper 1983; Cook \&
Brunt 1983; Wenzel 1984; Berriman et~al. 1986; and Watts et~al. 1986).
We will thereafter refer to this as the `low' or `minimum' brightness
state of V2051~Oph to differentiate it from the normal, quiescent state.
Thus, our data yield a rare (and lucky) opportunity to cleanly see the
white dwarf and bright spot in V2051~Oph, with reduced contributions
from the accretion disc and boundary layer to the total light.

Figure\,\ref{fig2} shows HST lightcurves at selected passbands,
wavelength range increasing upwards.  The eclipse shape is reminiscent
of those of the dwarf novae OY~Car and Z~Cha in quiescence (e.g., Wood
et~al.  1986, 1989), with two compact sources being eclipsed in sequence.
Sharp steps corresponding to the egress of the white dwarf and bright spot
are clearly seen in the G400H lightcurves.  A substantial fraction of
the emitted light comes from the white dwarf, which contributes $\simeq
60$ per cent of the total light at 3900--4300 \AA.  The G160L order zero
lightcurve shows the sequence of ingress of two compact sources. The
phases of ingress of the first eclipsed source are coincident with
those of the white dwarf as estimated from the ground-based data
(Fig\,\ref{fig3}). Hence, we can unambiguously associate the first
compact source with the white dwarf and, therefore, the source which
is eclipsed later is the bright spot.  Egress features of both sources
can also be easily identified in the order zero lightcurve. The flux of
the bright spot is larger at ingress than at egress. This is consistent
with the overall decrease in brightness along the run and is usually
understood in terms of anisotropic emission due to the foreshortening
of a spot `painted' at the edge of an optically thick accretion disc.
White dwarf and bright spot features are less well defined in the other,
lower signal-to-noise ratio G160L lightcurves.  Flickering activity --
usually higher in the ultraviolet than at optical wavelengths in CVs --
is considerably less impressive than observed in quiescence
(e.g, Warner \& O'Donoghue 1987), probably reflecting the lower
brightness state of the system at the epoch of our observations.
%

The HST dataset gives a clear and simultaneous measurement of the white
dwarf and bright spot eclipse phases, which will allow us to estimate
the binary mass ratio in section\,\ref{param}.

Figure\,\ref{fig3} display the ground-based observations.  These
lightcurves clearly show the total eclipse of a compact source at disc
centre, the white dwarf, and reveal that the white dwarf contributes
80--90 per cent of the light at that epoch, with very little evidence
of an accretion disc or bright spot.  Flickering activity is mostly
suppressed, which is possibly an indication that mass transfer was
substantially reduced at that epoch. The duration of the ingress/egress
feature ($\simeq 30$~s or 0.006 cycles, see section\,\ref{param})
is shorter than that in quiescence (Warner \& Cropper 1983; Berriman
et~al. 1986; Watts et~al. 1986) by a factor of two, possibly an
indication that, at those epochs, one was seeing a bright (and larger)
boundary layer around the white dwarf instead of the white dwarf itself.
This possibility was previously mentioned by Warner \& Cropper (1983).
%

The ground-based dataset yields an unambiguous measurement of the total
width of the eclipse of the white dwarf as well as of the duration of its
ingress/egress feature and will be crucial for our derivation of the
binary parameters in section\,\ref{param}.

\section{The binary parameters} \label{param}

In this section we use measurements of the contact phases of the white
dwarf and of the bright spot to develop a model for the binary
in V2051\,Oph based solely on quantities obtained from the photometry.

\subsection{Measuring contact phases} \label{contact}

\begin{figure*}
 \centerline{\psfig{figure=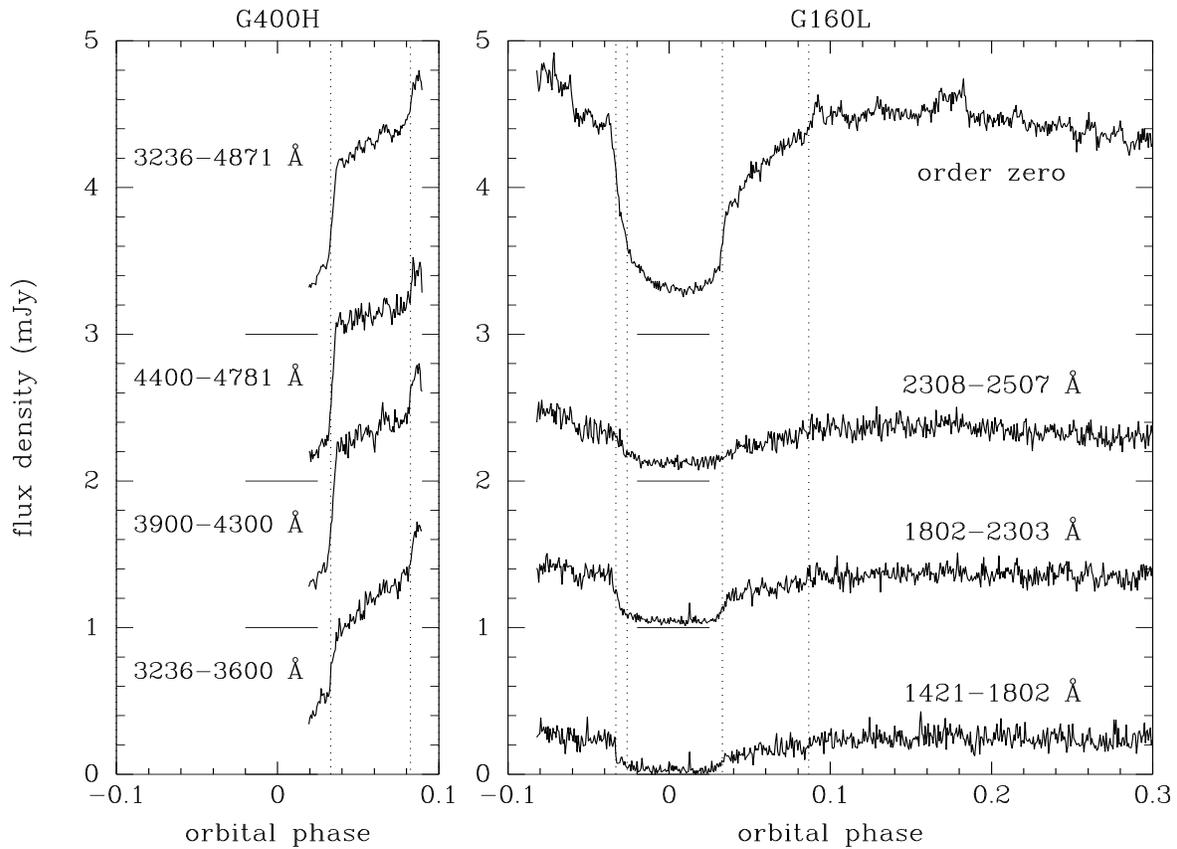,angle=-90,width=18cm,rheight=15cm}}
 \caption{HST lightcurves of V2051 Oph at selected passbands, for the
 G400H (left) and G160L (right) data. The 1421-1802 \AA\ lightcurves
 excludes the region of the C\,IV $\lambda 1550$ doublet.
 The curves are progressively displaced upwards by 1~mJy.
 Horizontal lines at mid-eclipse show the true zero level in each case.
 Vertical dotted lines mark ingress/egress phases of the white dwarf and
 bright spot as measured in section\,\ref{param}. }
 \label{fig2}
\end{figure*}

\begin{figure*}
 \centerline{\psfig{figure=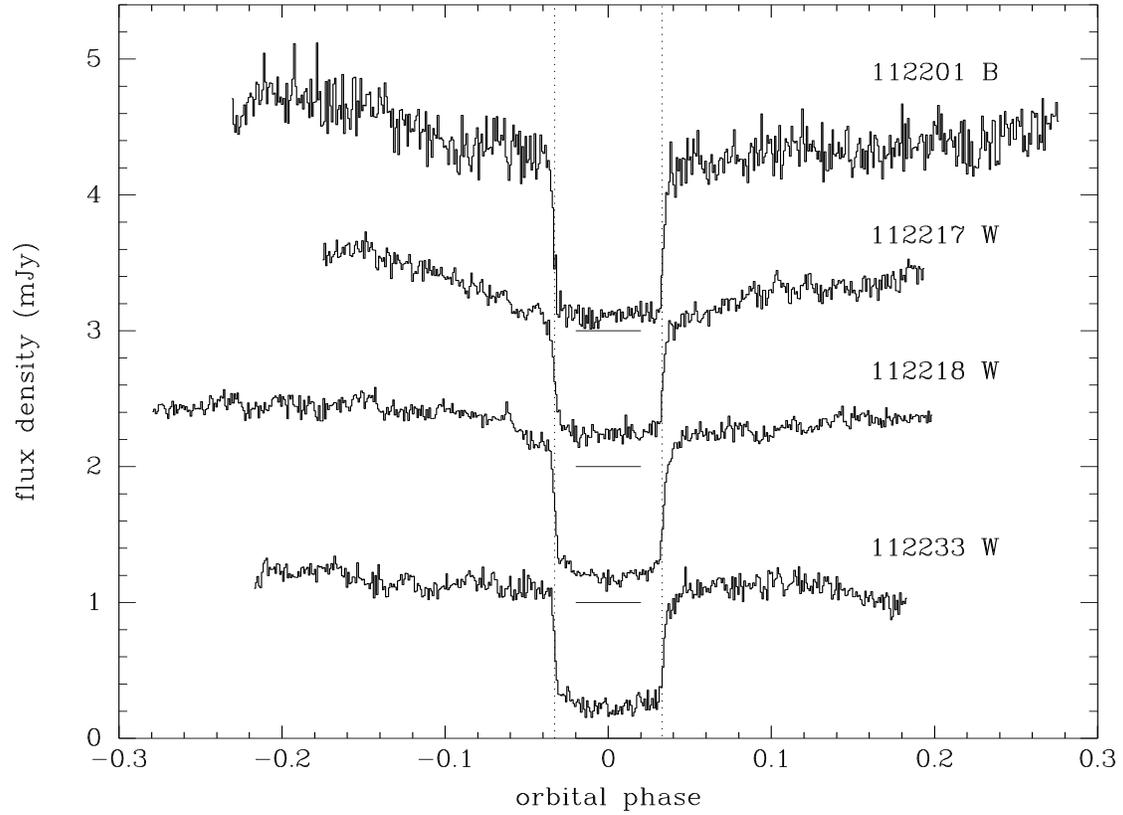,angle=-90,width=18cm,rheight=15cm}}
 \caption{Ground-based lightcurves of V2051 Oph.
 The white light lightcurves were artificially scaled to the same
 out-of-eclipse level of the B-band lightcurve for simplicity of display.
 The curves are progressively displaced upwards by 1~mJy.
 Horizontal lines at mid-eclipse show the true zero level in each case.
 Vertical dotted lines mark ingress/egress phases of the white dwarf as
 measured in section\,\ref{param}. }
 \label{fig3}
\end{figure*}

The ingress/egress phases of the occultation of the white dwarf (hereafter
WD) and of the bright spot (BS) by the secondary star provide information
about the geometry of the binary system and the relative sizes of
these components (Wood et al. 1986, 1989).  The contact phases can be
identified as rapid changes in the slope of the lightcurve, visible in
the lightcurves shown in Figs.~\ref{fig2} and \ref{fig3}.  We follow the
notation of Baptista et al. (1989), taking $\phi_{\rm w1}, \phi_{\rm w2}$
as those phases during which WD disappears behind the secondary star
and $\phi_{\rm w3}, \phi_{\rm w4}$ as the phases corresponding to the
reappearance of WD from eclipse.  We similarly define a set of phases
$\phi_{\rm b1-4}$ for BS.

The contact phases were measured with the aid of a cursor on a graphic
display of a median filtered version of the lightcurve.  The estimated
error for this procedure is 0.0006\,cycles.  Mid-ingress (egress) phases
for WD and BS ($\phi_{wi}, \phi_{we}$ and $\phi_{bi}, \phi_{be}$) were
computed as the mid point of the ingress (egress) feature and also as
the phase at which half of the compact source light is eclipsed.

\begin{figure*}
 \centerline{\psfig{figure=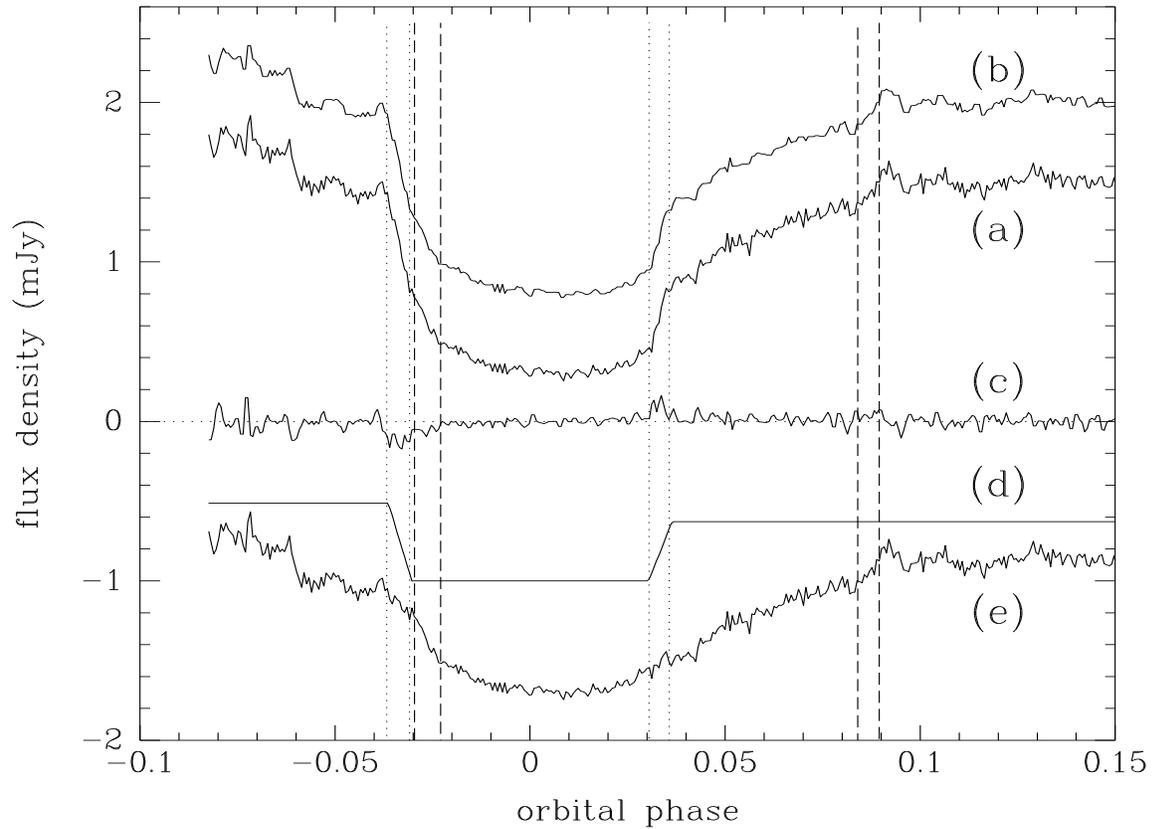,angle=-90,width=18cm,rheight=15cm}}
 \caption{Example of measuring contact phases in a lightcurve of V2051 Oph.
 (a) Original G160L white light lightcurve at a time resolution of $5 \times
 10^{-5}$ cycles. (b) median-filtered lightcurve, shifted upwards by 0.5 mJy.
 (c) the derivative of lightcurve (b) after removing the slowly-varying disc
 component, multiplied by a factor 2. (d) the reconstructed white dwarf
 lightcurve, shifted downwards by 1~mJy. (e) the lightcurve without the white
 dwarf component, shifted downwards by 2~mJy. Dotted lines mark the contact
 phases of the white dwarf and dashed lines mark the contact phases of the
 bright spot.  }
 \label{fig4}
\end{figure*}

We also employed the technique described by Wood, Irwin \& Pringle
(1985) to measure the contact points and to estimate the phases of
mid-ingress (egress), here defined as the points of minimum (maximum)
derivative in the lightcurve.  Figure\,\ref{fig4} illustrates the
procedure for the G160L order zero lightcurve.  The original lightcurve
[Fig.\,\ref{fig4}(a)] is smoothed with a median filter of width 0.0019
cycles (3 data points) [Fig.\,\ref{fig4}(b)] and its numerical derivative
is calculated.  The amount of filtering used at this point is a compromise
between the aim to suppress noise in the lightcurve as much as possible
and the need to preserve any real structures in the eclipse shape.
The ingress/egress of WD and BS can be seen as those intervals for which
the derivative curve is significantly different from zero.  The width
at half-peak intensity of these features yields a preliminary estimate
of their duration.  A spline function is fitted to the remaining regions
in the derivative curve to remove the contribution from the extended and
slowly varying eclipse of the disc. The spline-subtracted derivative curve
[Fig.\,\ref{fig4}(c)] is then analyzed by an algorithm which identifies
the points of extrema (the mid-ingress\,/\,egress phases) and the points
where the derivative starts to depart from the zero-constant level
(the contact points).  A median filter of equal width to that used in
the first part of the reduction is applied to the derivative curve to
aid in the detection.  A similar sequence is used to measure the contact
points and mid phases of BS.
%

\begin{figure*}
 \centerline{\psfig{figure=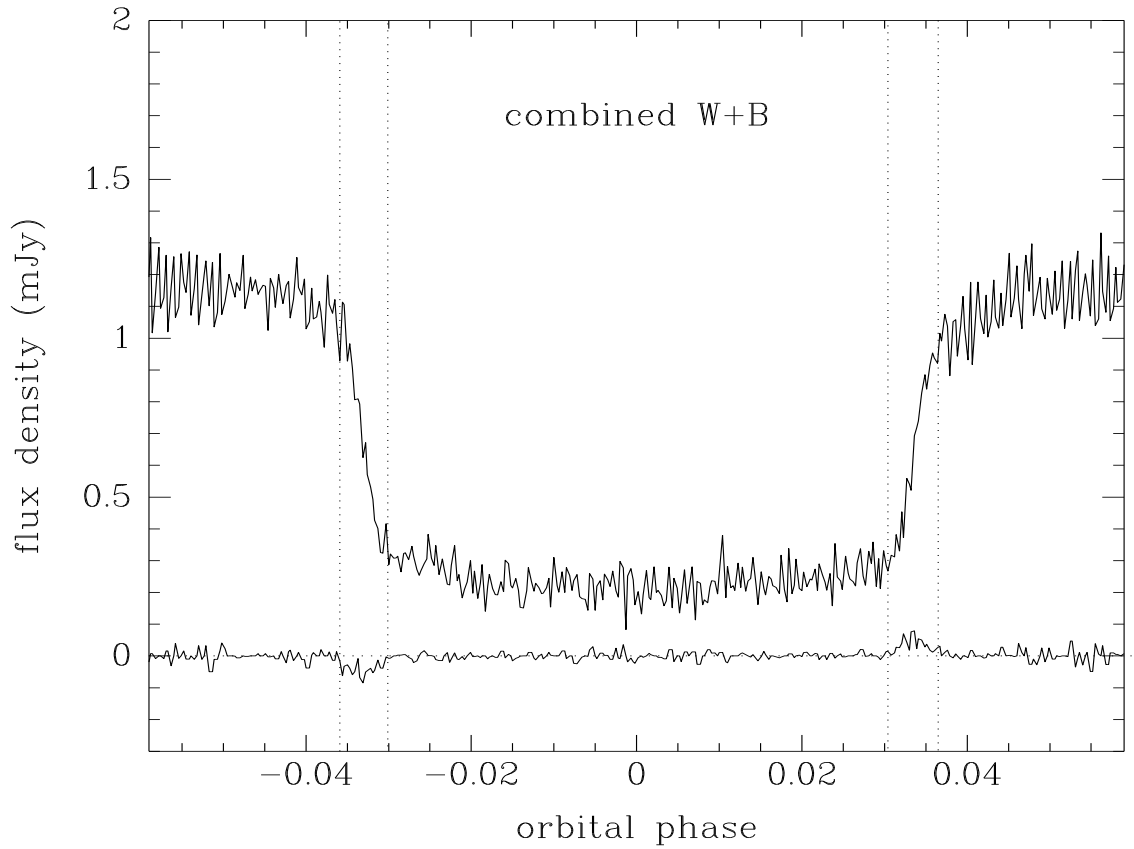,angle=-90,width=18cm,rheight=15cm}}
 \caption{The combined W+B lightcurve (runs L1 to L4) and its derivative curve.
 Vertical dotted lines depict the median contact phases of the WD as listed in
 Table\,\ref{wd-phases}. }
 \label{fig5}
\end{figure*}

Table~\ref{wd-phases} collects the measured contact phases, mid-ingress
and mid-egress phases of WD. Quoted values are the average of the
determinations from all methods described above. Uncertain measurements
are marked by a colon. Fig.\,\ref{fig5} depicts the median contact phases
of WD listed in Table\,\ref{wd-phases} in an enlarged view of the
combined W+B lightcurve and its derivative curve.
%
\begin{table*}
 \centering
 \begin{minipage}{140mm}
  \caption{Contact phases of the white dwarf.} \label{wd-phases}
  \begin{tabular}{@{}cccccccc@{}}

run & spectral range & $\phi_{w1}$ & $\phi_{w2}$ & $\phi_{w3}$ & $\phi_{w4}$
& $\phi_{wi}$ & $\phi_{we}$ \\ [1ex]

H1 & 3236-3600 \AA	& -- & -- & +0.0312 & +0.0369 & -- & +0.0339 \\
   & 3900-4300 \AA	& -- & -- & +0.0303 & +0.0365 & -- & +0.0335 \\
   & 4400-4781 \AA	& -- & -- & +0.0309 & +0.0363 & -- & +0.0337 \\
   & 3236-4781 \AA	& -- & -- & +0.0306 & +0.0361 & -- & +0.0337 \\ [0.7ex]
H2 & 1400-1800 \AA	& $-0.0348$ & ~$-0.0317$: & +0.0285 & +0.0347 & $-0.0332$ & +0.0331 \\
   & 1802-2303 \AA	& $-0.0361$ & $-0.0298$ & +0.0279 & +0.0361 & $-0.0330$ & +0.0325 \\
   & 1152-2506 \AA	& $-0.0359$ & $-0.0303$ & +0.0280 & +0.0341 & $-0.0328$ & +0.0334 \\
   &  order zero	& $-0.0366$ & $-0.0309$ & +0.0305 & +0.0355 & $-0.0334$ & +0.0330 \\
[0.7ex]
L1 &	B			& $-0.0374$ & $-0.0309$ & +0.0307 & +0.0367 & $-0.0332$ & +0.0332 \\
L2 &	W 			& $-0.0359$ & $-0.0285$ & +0.0308 & +0.0369 & $-0.0322$ & +0.0336 \\
L3 &	W 			& $-0.0356$ & $-0.0300$ & +0.0302 & +0.0367 & $-0.0333$ & +0.0334 \\
L4 &	W 			& $-0.0354$ & $-0.0303$ & +0.0304 & +0.0369 & $-0.0328$ & +0.0332 \\
combined W+B &			& $-0.0355$ & $-0.0298$ & +0.0306 & +0.0372 & $-0.0328$ & +0.0334 \\
[1ex]
median &				& $-0.0359$ & $-0.0301$ & +0.0304 & +0.0364 & $-0.0330$ & +0.0334 \\
error && $\pm0.0004$ & $\pm0.0005$ & $\pm0.0003$ & $\pm0.0003$ & $\pm0.0002$ & $\pm0.0002$
\end{tabular}
\end{minipage}
\end{table*}
%
%

Integrating the flux in the derivative curve Fig.\,\ref{fig4}(c) between
the first and second (third and fourth) contact phases we obtain estimates
of the WD flux at ingress (egress). The lightcurve of the WD is then
reconstructed by assuming that the flux is zero between ingress and
egress and that it is constant outside eclipse.  The reconstructed WD
lightcurve can be seen in Fig.\,\ref{fig4}(d), and the lightcurve
after removal of the WD component is shown in Fig.\,\ref{fig4}(e).

For the HST order zero lightcurve (fig.\,\ref{fig4}), the estimated WD
flux at ingress is $0.40\pm 0.05$ mJy, larger than the flux of $0.30
\pm 0.05$ mJy obtained from the egress feature.
This could be due to a phase dependent absorption of the
WD light, the absorption being more pronounced at egress phases.
The analysis of Catal\'an et~al. (1998) suggests that the WD in
V2051~Oph is veiled by a large number of blended Fe\,II lines (the so
called Fe\,II curtain).  This effect is reminiscent of that observed
previously in OY Car (Horne et~al.  1994) and has been attributed
to absorption by circumstellar material, possibly in the outer disc.
The observed difference in WD flux at ingress and at egress could be
associated with this effect if the absorption by the Fe\,II curtain is
more pronounced along the line of sight at egress phases.  This seems
to be the case in the novalike UX~UMa, whose ultraviolet eclipse mapping
reveals that the absorption by the Fe\,II curtain in the disc side closest
to the secondary star is more pronounced in the leading quadrant than
in the trailing quadrant (which contains the bright spot and gas stream)
and that this effect is probably responsible for a substantial reduction
in flux level from before to after eclipse (Baptista et~al. 1998b).

Measured BS contact phases, mid-ingress and mid-egress phases are
listed in Table~\ref{bs-phases}. As before, quoted values are the
average of the determinations from all methods described above.
For run H2, only values for the order zero lightcurve are listed since
the lightcurves at the other passbands (Fig.\,\ref{fig2}) do not yield
reliable measurements of BS phases.  The ingress phases $\phi_{b1}$,
$\phi_{b2}$ and $\phi_{bi}$ were measured from the WD subtrated
lightcurve (Fig.\,\ref{fig4}(e)), which provides an unblended, clean
view of the BS ingress feature. The short run H1 gives only the egress
phase of the BS, which occurs perceptibly earlier than at run H2.
Warner \& Cropper (1983) and Warner \& O'Donoghue (1987) have shown
that the eclipse shape of V2051~Oph (and therefore its disc structure)
is very variable and that significant changes occur on the timescale
of one orbit. One expects that similar changes in disc radius and BS
position do occur in the same timescale.  We have thus been cautious
not to combine egress phases from different cycles and, therefore,
for our procedure to determine the mass ratio $q$ (section\,\ref{geom})
we will use only the pair of mid-ingress/egress phases of BS measured
from run H2.  The measurements of mid-ingress/egress phases from the
three different methods have dispersions of 0.0005 cycles. However,
since only one light curve was used to estimate these values, we adopted
more conservative error bars of 0.001 cycles for the mid-ingress phase
and 0.002 cycles for the mid-egress phase.
%
\begin{table*}
 \centering
 \begin{minipage}{120mm}
  \caption{Contact phases of the bright spot.} \label{bs-phases}
  \begin{tabular}{@{}cccccccc@{}}

run & spectral range & $\phi_{b1}$ & $\phi_{b2}$ & $\phi_{b3}$ & $\phi_{b4}$
& $\phi_{bi}$ & $\phi_{be}$ \\ [1ex]

H1 & 3236-3600 	& -- & -- & +0.0811 & +0.0855 & -- & +0.0831 \\
   & 3900-4300 	& -- & -- & +0.0804 & +0.0848 & -- & +0.0822 \\
   & 4400-4781 	& -- & -- & +0.0811 & +0.0836 & -- & +0.0825 \\
   & 3236-4781 	& -- & -- & +0.0809 & +0.0831 & -- & +0.0821 \\ [0.7ex]
H2 & order zero & $-0.0295$ & $-0.0226$ & +0.0841 & +0.0892 & $-0.0260$ & +0.0866 \\ [1ex]
\end{tabular}
\end{minipage}
\end{table*}

The duration of the WD eclipse is defined as
\begin{equation}
\Delta\phi= \phi_{we} - \phi_{wi} \; ,
\end{equation}
and the centre of the WD eclipse (the inferior conjunction of the
binary) is written as
\begin{equation}
\phi_0 = 1/2\:(\phi_{we} + \phi_{wi}) \; .
\end{equation}
These quantities are collected in Table~\ref{wd-param}.
The median of the measurements from all lightcurves yields
$\Delta\phi = 0.0662 \pm 0.0002$ cycles, where the quoted error is
the median of the absolute deviations with respect to the median.
Similarly, we have $\phi_0 = +0.0000 \pm 0.0002$ cycles, which indicates
that the centre of the WD eclipse corresponds to phase zero.

The difference between the first and second (third and fourth)
WD contact phases yield the phase width of the WD ingress (egress),
$\Delta_{\rm wi}$ ($\Delta_{\rm we}$).  These quantities are also listed
in Table~\ref{wd-param}.  A median from all values of $\Delta_{\rm wi}$
and $\Delta_{\rm we}$ yields $\Delta_{\rm wd}= 0.0060 \pm 0.0005$\,cycles,
where again the error is the median of the absolute deviations with
respect to the median.
%
\begin{table*}
 \centering
 \begin{minipage}{90mm}
  \caption{White dwarf eclipse parameters.} \label{wd-param}
  \begin{tabular}{@{}cccccc@{}}
run & spectral range & $\Delta\phi$ & $\phi_0$ & $\Delta_{\rm wi}$ & $\Delta_{\rm we}$ \\
[1ex]
H1 		& 3236-3600 	&  &  &  & 0.0057 \\
   		& 3900-4300 	&  &  &  & 0.0062 \\
   		& 4400-4781 	&  &  &  & 0.0054 \\
   		& 3236-4781 	&  &  &  & 0.0056 \\ [0.7ex]
H2 		& 1400-1800 	& 0.0663 & $-0.0001$ & ~0.0032: & 0.0062 \\
       	& 1802-2303 	& 0.0655 & $-0.0003$ & 0.0063  & 0.0082 \\
       	& 901-2506 		& 0.0662 &  +0.0003  & 0.0056  & 0.0061 \\
       	& order zero 	& 0.0664 & $-0.0002$ & 0.0057  & 0.0050 \\ [0.7ex]
L1		& B 			& 0.0664 &  +0.0000  & 0.0065  & 0.0060 \\
L2 		& W				& 0.0658 &  +0.0007  & 0.0074  & 0.0061 \\
L3 		& W 			& 0.0667 &  +0.0000  & 0.0056  & 0.0065 \\
L4		& W 			& 0.0660 &  +0.0002  & 0.0051  & 0.0065 \\
combined W+B &		& 0.0662 &  +0.0003  & 0.0057  & 0.0066 \\ [1ex]
median &			& 0.0662 &  +0.0000 & ~0.0057 & ~0.0061 \\
error  &			& $\pm 0.0002$~ & $\pm 0.0002$ & $\pm 0.0006$ & $\pm 0.0007$ \\
\end{tabular}
\end{minipage}
\end{table*}

\subsection{Mass ratio, inclination and disc radius} \label{geom}

Making the usual assumption that the secondary star fills its Roche
lobe and given the duration of the eclipse of the central parts of
the disc, $\Delta\phi$, there is a unique relation between the mass
ratio $q= M_2/M_1$ and the binary inclination $i$ (Bailey 1979;
Horne 1985). From Table\,\ref{wd-param}, the width of the eclipse in
V2051~Oph is $\Delta\phi= 0.0662$. This gives the constraint $q>0.116$,
with $q=0.116$ if $i=90\degr$.

When combined with the measured contact phases of the WD and BS, this
relation gives a unique solution for $q$, $i$, and $R_{bs}/R_{L1}$,
where $R_{bs}$ is the distance from disc centre to the BS (usually taken
to be the disc radius) and $R_{L1}$ is the distance from disc centre
to the inner lagrangian point L1.  The following assumptions are made:
(i) the infalling gas stream leaving the L1 point describes a balistic
trajectory in the primary lobe (Lubow \& Shu 1975), and (ii) the bright
spot is located at the position at which the stream hits the edge of
the disc.  The theoretical trajectories described by a test particle in
the primary lobe are solely determined by the mass ratio of the binary.
For increasing $q$ values the trajectories are closer to the line joining
both stars, as the angular momentum of the infalling gas stream after
leaving the L1 point is progressively smaller. The correct mass ratio, and
hence inclination, are those for which the calculated stream trajectory
passes through the observed position of the bright spot.  This method
was first developed by Smak (1971) and has since then been applied to
various CVs (e.g., Fabian et~al. 1978; Cook \& Warner 1984;
Wood et~al. 1986, 1989; Baptista, Steiner \& Cieslinski 1994).

\begin{figure*}
 \centerline{\psfig{figure=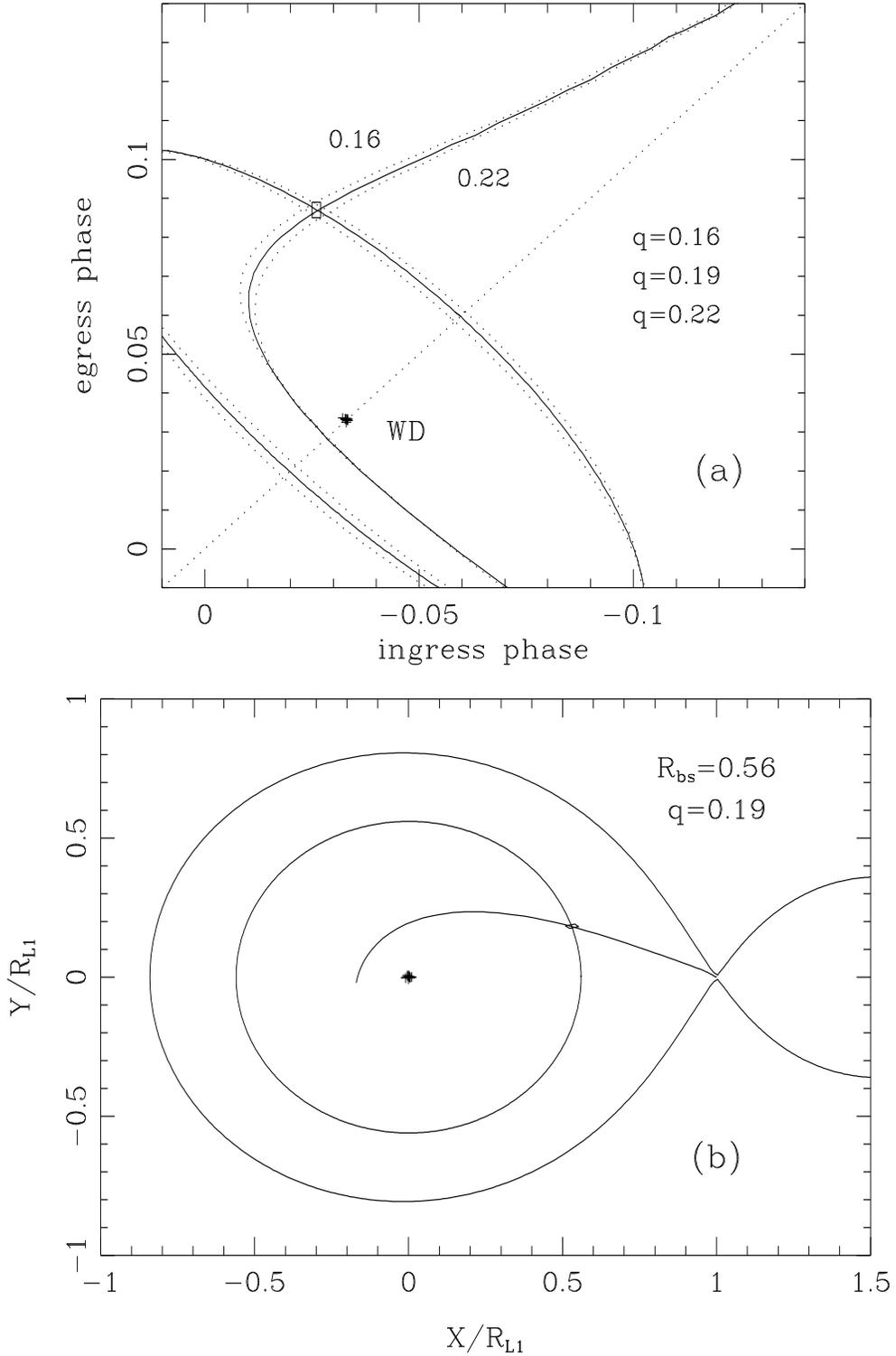,width=16.5cm,rheight=21cm}}
  \caption{Inferring the binary geometry from the ingress/egress phases
  of WD and BS. (a) Ingress-egress phases diagram. The observed phases of
  mid-ingress/egress of WD are marked with crosses, those of BS with a small
  box. A diagonal dotted line depict the line joining the component stars.
  Theoretical gas stream trajectories for various values of $q$ are plotted.
  The stream of matter passes through the position of BS for $q=0.19$.
  The squashed circle represents the accretion disc whose edge passes
  through the position of BS, for $q=0.19 \pm 0.03$.
  This yields a disc radius of $R_{bs}= 0.56 \; R_{L1}$.
  (b) the adopted geometry of the binary for $q=0.19$.
  The observed positions of WD and BS are shown with the theoretical gas
  stream and a disc of radius $R_{bs}= 0.56 \; R_{L1}$. }
 \label{fig6}
\end{figure*}

Figure\,\ref{fig6} illustrates the determination of $q$, $i$ and
$R_{bs}/R_{L1}$ from the eclipse phases.  Fig.\,\ref{fig6}(a) shows a
diagram of ingress versus egress phases for the measurements of the WD and
BS in Tables\,\ref{wd-phases} and \ref{bs-phases}. Measurements of the WD
ingress and egress are shown as the cluster of small crosses around phases
($-0.033, +0.033$) in the lower portion of the diagram. The small box
in the upper left corner marks the 1-$\sigma$ range of the mid-ingress
and mid-egress phases of the BS.  Theoretical gas stream trajectories
corresponding to a set of pairs $(i,q)$ are also shown. The trajectories
were computed by solving the equations of motion in a coordinate system
synchronously rotating with the binary, using a 4th order Runge-Kutta
algorithm (Press et al. 1986) and conserving the Jacobi integral constant
to one part in $10^{6}$.
%

The theoretical trajectory that passes through the position of
the BS has $q= 0.19 \pm 0.03$ and $i= 83\fdg 3 \pm 1\fdg 4$, where the
uncertainties are taken from the standard deviation of the points about
the trajectory of best fit.  Fig.\,\ref{fig6}(b) shows the geometry of
the binary system for $q=0.19$. For this mass ratio, the relative size
of the primary Roche lobe is $R_{L1}/a= 0.66$, where $a$ is the
orbital separation.

The squashed circle in Fig.\,\ref{fig6}(a) represents the accretion
disc whose edge passes through the position of BS for the adopted mass
ratio.  This corresponds to a disc radius of $R _{bs}/R_{L1}= (0.56 \pm
0.02)$.  A circle with this radius is depicted in Fig.\,\ref{fig6}(b).
The calculated radius is half way between the radius expected for
zero-viscosity discs, $R_d/R_{L1}\!=\!0.27$ (Flannery 1975), and that
expected for pressureless discs, $R_d/R_{L1}\!=\!0.76$ (Paczy\'{n}ski
1977).

\subsection{Masses and radii of the component stars}

The duration of the ingress\,/\,egress of WD yields an estimate
of the relative size of the white dwarf.
This information can be used together with the calculated
mass ratio to constraint the masses and radii of the component stars.
We use the approximate relations of Ritter (1980),
\begin{equation}
2 \pi \Delta_{\rm wd} \simeq 2\:{R_1}/[z(q)\,a\,\sin \theta]\; ,
\label{wd.raio}
\end{equation}
and
\begin{equation}
\cos \theta = \frac{a}{R_{2}}\,\cos\,i \; ,
\end{equation}
where $R_1$ and $R_2$ are the radii of the WD and of the secondary
star, and $z(q)$ is the relative distance from disc centre to the point
tangent to the surface of the secondary that marks the beginning/end
of the eclipse of WD (Baptista et al. 1989).  The value of $z(q)$ is a
by-product of the inclination-determination routine (section\,\ref{geom})
and is usually close to unity.  In our case, for $q=0.19$ we have $z=0.96$.
The error introduced by the use of these expressions is of order of
$(R_1/R_2)^2$ and for the parameters listed in Table\,\ref{param}
is smaller than 0.5 per cent.

Combining Eq.\,(\ref{wd.raio}) with Kepler's third law we can write
an expression relating the mass and the radius of the white dwarf for
V2051~Oph,
\begin{equation}
R_1/R_\odot = 2.349\:\:\Delta_{\rm wd} \: (M_1/M_{\rm ch})^{1/3} f(q) \; ,
\label{uu.rwd}
\end{equation}
where $M_{\rm ch}\!=\!1.44\: M_\odot$ is the Chandrasekhar mass limit
and $f(q) = z\,\sin\theta \: (1+q)^{1/3}$.  Another relation between
$R_1$ and $M_1$ can be obtained from Nauenberg's (1972) analytical
approximation to the mass-radius relation for cool degenerate white
dwarfs of Hamada \& Salpeter (1961),
\begin{equation}
R_1/R_\odot = 0.0112 \left[ \left( \frac{M_1}{M_{\rm ch}}
\right)^{-2/3}\!\!\!\!- \left( \frac{M_1}{M_{\rm ch}} \right)^{2/3}
\right]^{1/2} \;\; .
\label{nauen}
\end{equation}
\noindent
Combining Eqs.\,(\ref{uu.rwd}) and (\ref{nauen}) one can solve for
$M_1(q,\Delta_{\rm wd})$ and the remaining system para\-meters.

\begin{figure*}
 \centerline{\psfig{figure=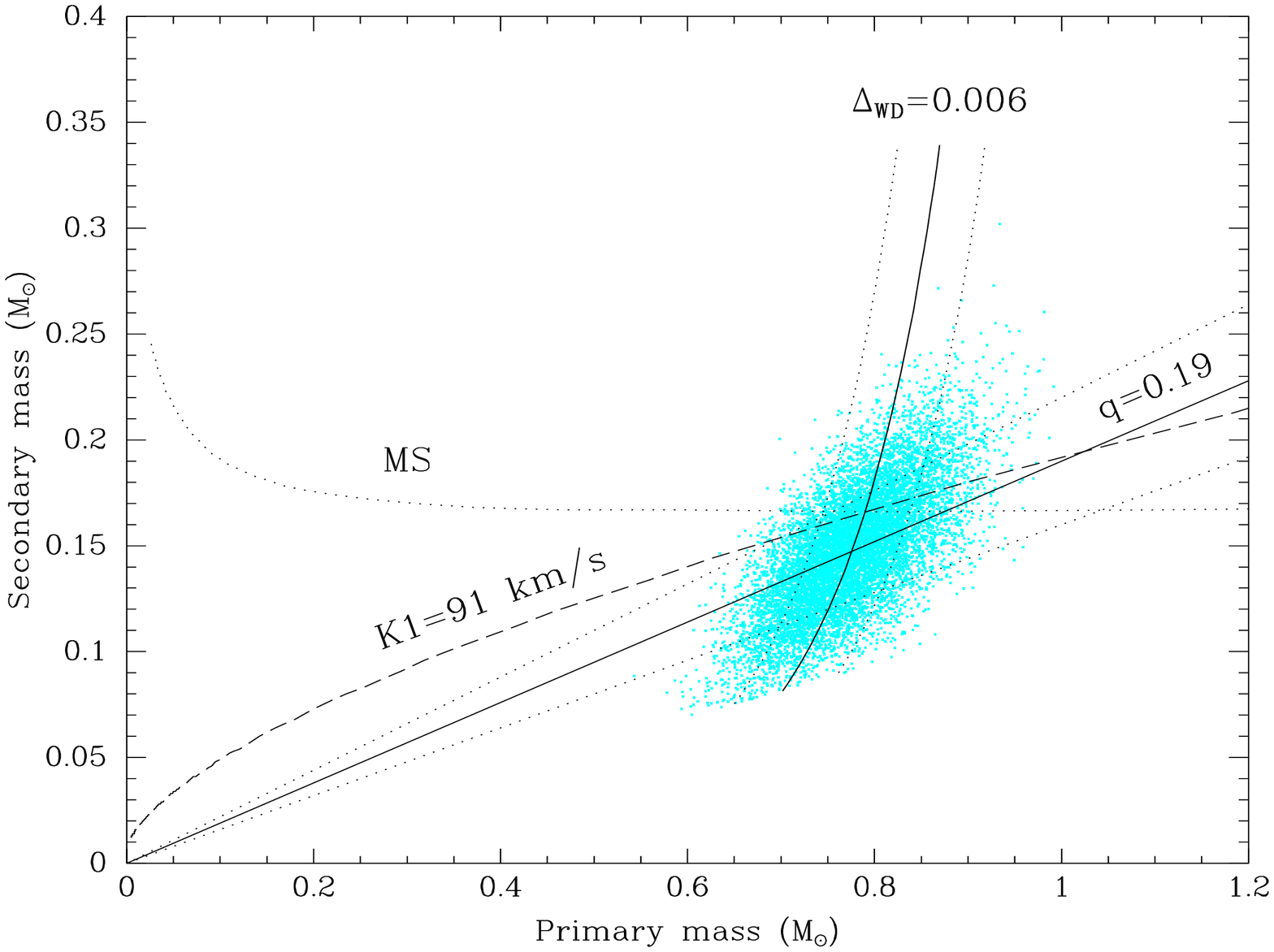,angle=-90,width=18cm,rheight=15cm}}
  \caption{Primary - Secondary mass diagram for V2051 Oph.
  Solid lines show the constraints obtained from the measured $\Delta_{wd}$
  and the inferred mass ratio of $q=0.19$. Dotted lines indicate the
  1-$\sigma$ limit on these relations. The mass function for a white dwarf
  radial velocity of $K_1=91$ km/s (Watts et~al. 1986) is shown as a dashed
  line and a mass-radius relation of $R_2/R_\odot = M_2/M_\odot$ is shown
  as a dotted line. The gray cloud of points shows the confidence region and
  is the result of a $10^4$ points Monte Carlo simulation with the values of
  $\Delta_{\rm wd}$ and $q$. }
\label{mass}
\end{figure*}

The primary-secondary mass diagram for V2051~Oph can be seen in
Fig.\,\ref{mass}. A Monte Carlo propagation code was used to estimate the
errors in the calculated parameters.  The values of the input parameters
$q$ and $\Delta_{\rm wd}$ are independently varied according to
Gaussian distributions with standard deviation equal to the corresponding
uncertainties.  The results, together with their 1-$\sigma$ errors, are
listed in Table\,\ref{sys_param}.  The cloud of points in Fig.\,\ref{mass}
was obtained from a set of $10^4$ trials using this code. The highest
concentration of points indicates the region of most probable solutions.
%

\begin{table}
 \centering
 \begin{minipage}{140mm}
  \caption{The calculated parameters of V2051 Oph.} \label{sys_param}
  \begin{tabular}{@{}llll@{}}
$q$		& $0.19 \pm 0.03$			& $a$/R$_\odot$		& $0.64\pm 0.02$ \\
$i$		& $83.3\degr \pm 1.4\degr$	& R$_{\rm L1}/a$	& $0.66\pm 0.01$ \\
M$_1$/M$_\odot$	& $0.78 \pm 0.06$	& R$_{\rm bs}$/R$_{\rm L1}$	& $0.56\pm 0.02$ \\
R$_1$/R$_\odot$	& $0.0103 \pm 0.0007$ & $\alpha_{\rm bs}$		& $19\degr\pm 1\degr$ \\
M$_2$/M$_\odot$	& $0.15 \pm 0.03$	& K$_1$ (km/s)		& $83 \pm 12$ \\
R$_2$/R$_\odot$	& $0.16 \pm 0.01$	& K$_2$	(km/s)		& $436 \pm 11$
\end{tabular}
\end{minipage}
\end{table}

\section{Discussion} \label{discussion}

\subsection {The reliability of the photometric model}

Our photometric model is based on the inferred $\Delta_{\rm wd}$
and $q$ values. Here we discuss systematic errors which may affect these
quantities and their influence on the derived parameters of V2051~Oph,
and compare the derived parameters with those obtained by other authors.

Errors in $q$ are related to errors in measuring mid-ingress/egress phases,
in particular the BS phases. WD eclipse phases were precisely measured
in the ground-based lightcurves, where the BS and the accretion
disc give negligible contribution to the total light. These phases
helped us to identify the WD eclipse features in run H2 without
ambiguities. The lightcurve after the subtraction of the WD contribution
(fig.\,\ref{fig4}(d)) clearly shows the asymmetric eclipse of the compact
BS and yielded accurate measurements of its mid-ingress/egress phases. The
measured BS phases map into a small region in the ingress/egress phases
diagram from which a precise value of $q$ was obtained. Therefore,
the mass ratio seems well constrained.

The remaining binary parameters were derived under the assumption that
the duration of the ingress/egress of the central object, $\Delta_{\rm
wd}$, gives a direct measure of the radius of the white dwarf. This
supposition may not always be correct.

It may be possible that the observed compact object of Fig.\,\ref{fig5}
is a hot and opaque boundary layer involving the white dwarf, in which
case the true white dwarf would be smaller than assumed (Wood \& Crawford
1986). In this case, the primary mass found is a lower limit.  However,
notice that increasing the white dwarf mass along the line $q=0.19$
in Fig\,\ref{mass} results in a secondary star which is overmassive in
comparison to a main sequence star of same radius.  This would be in
marked contrast with the other short-period eclipsing CVs, where the
results seem to point in the opposite direction, i.e., secondary stars
which are undermassive in comparison to main sequence stars of same radius
(Wood et~al. 1986, 1989; Horne, Wood \& Stiening 1991).
Other effects that may result in an overestimated white dwarf radius
include the questionable application of a mass-radius relation for cool
white dwarfs to a relatively hot object (Horne, Wood \& Stiening 1991),
and a possible spherical distortion for a white dwarf rotating close
to breakup velocity (Wood \& Horne 1990). However, the corrections in
the derived parameters implied by these effects ($\simlt 5$ per cent)
are smaller than the uncertainty in the value of $\Delta_{wd}$ and were
not taken into consideration.

Alternatively, could the white dwarf be larger than inferred from our
value of $\Delta_{wd}$? We now address two possibilities in this regard.

Previous eclipse photometry of V2051~Oph indicate a duration of
the ingress/egress of the compact central source of $\Delta_{wd}
\simeq 0.009-0.012$ (Warner \& Cropper 1983; Cook \& Brunt 1983;
Watts et~al. 1986), substantially larger than measured from our
observations. However, the variability and the clear correlation between
brightness and duration of ingress/egress of this source led Warner
\& Cropper (1983) to suggest that it might be a bright (and variable)
boundary layer around the white dwarf and not the white dwarf itself.
This scenario is supported by our observations at minimum light, when
mass accretion has probably been substantially reduced and the boundary
layer has accordingly dimmed, leaving a bare, fainter, and much smaller
white dwarf as the only light source at disc centre.

Warner \& O'Donoghue (1987) suggested that V2051 Oph could be a low
field polar. Following their suggestion, we explored the possibility
that the observed compact source at disc centre is in fact a bright
spot at the surface of the white dwarf, corresponding to the pole of an
accretion column.  If the white dwarf rotates synchronously with the
binary, one would expect the lightcurves to show a strong orbital
modulation due to the self occultation of the accretion column.  For
example, at phase $\phi=0.25$ about half of the accretion column would
have disappeared behind the white dwarf resulting in a reduction of
$\simgt 40$ per cent in flux of the lightcurve.  On the other hand, if
the white dwarf is non synchronously rotating, one would expect to see
two conspicuous effects in the lightcurves: (1) large amplitude
($\simeq 40$ per cent) pulsations with the rotation period of the white
dwarf, and (2) measurable wanderings of the ingress/egress phases and
eclipse width of the compact source at disc centre as the accretion
column is seen at different aspects.  Our observations clearly exclude
both possibilities: there is no sign of strong orbital modulation, no
sign of pulsations, and no detected changes in eclipse phases.
Therefore, we are left with the conclusion that the compact source at
disc centre of our observations is indeed the white dwarf.

It is worth mentioning that our observations do not discard the
possibility that V2051~Oph possess a magnetic white dwarf.  If the
observed minimum brightness state corresponds to a phase of reduced mass
transfer and negligible or no mass accretion, then no emission would be
produced in the accretion column, which could therefore remain undetected.

We now turn to the comparison of our derived parameters with those
obtained by Watts et~al. (1986).
Their dynamical model of V2051~Oph is based on spectroscopy of the
accretion disc's emission lines.  Radial velocity for the H$\beta$ and
H$\gamma$ lines yields a $K_1= 111\; {\rm km\; s}^{-1}$ and a phase lag
of $+0.13$ cycles with respect to the photometric phase.  Phase lags of
$0.1 - 0.2$ cycles between photometric and spectroscopic conjunction
are seen in many of the CVs for which emission lines radial velocity
curves were measured.  Asymmetries in the line intensity distribution
(which displace the light centroid from disc centre) or the existence
of a non-circular component to the velocity of the emitting gas (as it
may happen in the presence of a disc wind) lead to systematic errors in
the amplitude and phase of the measured $K_1$. They show that the
observed phase lag of V2051~Oph can be explained if the accretion disc
is asymmetric in the emission lines, with the trailing lune being
$\simeq 60$ per cent brighter than the leading lune. The measured $K_1$
will be overestimated by $\simeq 22$ per cent in this case.  The
correction of their $K_1$ value by this effect eliminates the phase lag
and brings their measurement to $K_1= 91\; {\rm km\; s}^{-1}$, into
good accordance to our prediction in Table\,\ref{param}.

Watts et~al. (1986) used the double peaked Balmer line profiles and
their behaviour during eclipse to constrain the mass ratio of the binary.
Their model predicts a larger mass ratio ($q=0.26\pm 0.04$) and a much
smaller white dwarf mass ($M_1= 0.43\pm 0.05\; M_\odot$) than ours.
We note that their fit to the line behaviour during eclipse seems
of rather poor quality.
The larger spectroscopic mass ratio predicts that the bright spot
eclipse occurs earlier than observed, with a phase difference at the
2-$\sigma$ level. The spectroscopic model also predicts a duration of
the ingress and egress of the white dwarf of $\Delta_{wd}= 0.011$,
which clearly conflicts with the observed duration of $\Delta_{wd}=
0.0060\pm 0.0005$.

It would be useful to have another observational constraint on the
binary parameters. The best way of checking the reliability of the
above assumptions in calculating binary parameters is to measure
the radial velocity of the secondary star, $K_2$.

\subsection {The low brightness state: reduced mass transfer rate?}

At the time of our observations V2051~Oph was in a much lower brightness
state than at quiescence, when it has $V\simeq 15$ mag. Since it was
observed to be in this lower brightness state both in 1996 January and in
1996 June, it may be possible that this is a relatively prolonged minimum
state of duration a few months.  Recovery of photometric observations
of V2051~Oph during this period would be important to clarify this point.

The system has been observed to go into outburst once, where it increased
in brightness to reach $V\simeq 13.2$  (Warner \& O'Donoghue 1987). If the
outburst is caused by the disc instability mechanism, than the disc is
{\em already} in a state of low viscosity, low mass accretion rate when
the system is at $V\simeq 15$ mag. Therefore, it is hard to explain the
observed low brightness state in terms of disc instability.  The simplest
way to understand the reduced brightness level of V2051~Oph is to consider
it the consequence of a significant reduction in the mass transfer rate.

\subsection {The secondary star mass and its evolutionary consequences}

Our photometric model indicates that V2051~Oph has a larger mass ratio
and a more massive white dwarf than the other eclipsing dwarf novae of
similar period. Probably more relevant, it has a secondary star which
is significantly more massive than those of Z~Cha, OY~Car and HT Cas.
Since the inferred size of the secondary in these systems is similar,
$R_2/R_\odot \simeq 0.15$, this means that the secondary star of
V2051~oph obeys a mass-radius relationship which predicts significantly
denser stars for a given mass.

Theoretical evolutionary models of CVs predict that the long term,
continuous mass loss brings the secondary stars out of thermal equilibrium
and make these stars larger than a main sequence star of same mass
(Rappaport, Joss \& Webbink 1982). Hence, the fact that the secondary
stars in Z~Cha, OY~Car and HT Cas are oversized in comparison to main
sequence stars of same mass suggests that these systems are relatively
old cataclysmic variables, for which mass transfer (and loss from the
secondary) has been occurring for some $10^9$ years. On the other hand, the
fact that the secondary star of V2051~Oph obeys a typical main sequence
mass-radius relationship suggests that V2051~Oph is a relatively young
system, for which mass loss has not occurred for long enough time to
bring the secondary star out of thermal equilibrium.  Further support to
this suggestion comes from the following argument.  If we assume that CVs
evolve by loosing mass in recurrent nova outbursts (at a rate of $\simeq
10^{-5}\; M_\odot$ at every $10^4 - 10^5$ years), than V2051~Oph
will need about $10^9$ years to reduce its total mass of about $0.9 \;
M_\odot$ to the typical total mass of $\simeq 0.7 \; M_\odot$
of the other, older dwarf novae of similar orbital period.

\section{Conclusions} \label{fim}

The analysis of HST/FOS and ground-based eclipse observations of the
dwarf novae V2051~Oph yielded the following results:

\begin{enumerate}

\item V2051~Oph was observed in a low brightness state, at $B \simeq
16.2$ mag. In comparison to the average IUE spectra, the ultraviolet
continuum and emission lines appear reduced by factors of,
respectively, $\simeq 4$ and $\simeq 5$.  Flickering activity is mostly
suppressed and the lightcurve shows the eclipse of a compact white dwarf
at disc centre which contributes $\simeq 60$ per cent of the total
light at 3900--4300 \AA. In 1996 January, the bright spot is clearly
seen in the lightcurve, allowing a precise measurement of its eclipse
phases. In 1996 June only the white dwarf is seen in the lightcurve
and its radius can be estimated from accurate measurement of the width
of its ingress/egress.  It is suggested that this state is the
consequence of a substantial reduction in mass transfer rate.

\item We developed a purely photometric model for the binary from the
eclipse phases of the white dwarf and bright spot.
For the gas stream trajectory to pass through the position of the
bright spot a mass ratio $q= 0.19 \pm 0.03$ is required.
The corresponding inclination is $i= 83\fdg 3 \pm 1\fdg 4$.

\item The duration of the ingress/egress of the white dwarf combined
with the Hamada-Salpeter mass-radius relationship and the obtained mass
ratio give $M_1= 0.78\pm 0.06 \; M_\odot$ and $M_2= 0.15 \pm 0.03 \;
M_\odot$. Calculated parameters are summarized in Table\,\ref{sys_param}.

\item The secondary of V2051~Oph is significantly more massive than the
secondaries of the other ultra-short period dwarf novae and seems to obey
a ZAMS mass-radius relationship.  V2051~Oph is probably a relatively
young system, whose secondary star had not enough time to evolve out
of thermal equilibrium. 

\item Our photometric model predicts $K_1= 83 \pm 12$ km s$^{-1}$, which is
in accordance with the velocity amplitude obtained from the emission lines
after a correction for asymmetric line emission in the disc is made
(Watts et~al. 1986).

\end{enumerate}

\section*{Acknowledgments}

We thank Andrew King and an anonymous referee for valuable comments and
suggestions on an earlier version of the manuscript.
RB acknowledges financial support from CNPq/Brazil through grant no. 300\,354/96-7.

\begin{thebibliography}{99}
\bibitem{b1} Bailey J., 1979, MNRAS, 187, 645
\bibitem{b2} Baptista R., Jablonski F.J., Steiner J.E., 1989, MNRAS, 241, 631
\bibitem{b3} Baptista R., Steiner J. E., Cieslinski D., 1994, ApJ, 433, 332
\bibitem{b4} Baptista R., Tripplet L., Bond H., 1998a, in preparation
\bibitem{b5} Baptista R., Horne K., Wade R., Hubeny I., Long K., Rutten R.G.M.,
		1998b, MNRAS, in press
\bibitem{b6} Berriman G., Kenyon S., Bailey J., 1986, MNRAS, 222, 871
\bibitem{b43} Bond H., Wagner R.L., 1977, IAU Circular 3049
\bibitem{b7} Bruch A., 1983, IBVS, 2287
\bibitem{b42} Catal\'an M.S., Horne K., Cheng F.-H., Marsh T., Hubeny I., 1998,
		preprint
\bibitem{b8} Cook M.C., Brunt C.C., 1983, MNRAS, 205, 465
\bibitem{b9} Cook M.C., Warner B., 1984, MNRAS, 207, 705
\bibitem{b10} Eracleous M., Horne K., 1994, ApJ, 433, 313
\bibitem{b11} Fabian A.C., Lin D.N.C., Papaloizou J., Pringle J.E., Whelan
		J.A.J., 1978, MNRAS, 184, 835
\bibitem{b12} Flannery B.P., 1975, MNRAS, 170, 325
\bibitem{b13} Hamada T., Salpeter E.E., 1961, ApJ, 134, 683
\bibitem{b14} Horne K., 1985, MNRAS, 213, 129
\bibitem{b15} Horne K., Marsh T.R., Cheng F.-H., Hubeny I., Lanz T., 1994,
		ApJ, 426, 294
\bibitem{b16} Horne K., Wood J.H., Stiening R.F., 1991, ApJ, 378, 271
\bibitem{b17} Graham J.A., 1982, PASP, 94, 244
\bibitem{b18} Jablonski F.J., Baptista R., Barroso Jr. J., Gneiding C.,
		Rodrigues F., Campos R.P., 1994, PASP, 106, 1172
\bibitem{b19} Lamla E., 1981, Landolt-B\"{o}rnstein - Numerical Data and
		Functional Relationships in Science and Technology, Vol.\,2, eds.
		K.\,Schaifers \& H.H.\,Voigt, Springer-Verlag
\bibitem{b20} Lubow S.H., Shu F.H., 1975, ApJ, 198, 383
\bibitem{b21} Nauenberg M., 1972, ApJ, 175, 417
\bibitem{b22} Paczy\'{n}ski B., 1977, ApJ, 216, 822
\bibitem{b23} Press W.H., Flannery B.P., Teukolsky S.A., Vetterling W.T.,
		1986, Numerical Recipes, Cambridge University Press
\bibitem{b24} Pringle J. E., Verbunt F., Wade R. A., 1986, MNRAS, 221, 169
\bibitem{b25} Rappaport S., Joss P.C., Webbink R.F., 1982, ApJ, 254, 616
\bibitem{b26} Robinson E.R., 1992, Vi\~{n}a del Mar Workshop on Cataclysmic
		Variable Stars, ASP Conference Series Vol. 29, Ed. N.\,Vogt, Astron.
		Soc. Pac., San Francisco, p.3
\bibitem{b27} Ritter H., 1980, A\&A, 86, 204
\bibitem{b28} Sanduleak N., 1972, IBVS, 663
\bibitem{b29} Smak J., 1971, Acta Astr., 21, 15
\bibitem{b30} Stone R.P.S., Baldwin J.A., 1983, MNRAS, 204, 347
\bibitem{b31} Vogt N., 1983, A\&AS, 53, 21
\bibitem{b41} Warner B., 1996, Cataclysmic Variable Stars, Cambridge
		Astrophysics Series 28, Cambridge University Press
\bibitem{b32} Warner B., Cropper M., 1983, MNRAS, 203, 909
\bibitem{b33} Warner B., O'Donoghue D., 1987, MNRAS, 224, 733
\bibitem{b34} Watts D.J., Bailey J., Hill P.W., Greenhill J.G., McCowage C.,
		Carty T., 1986, A\&A, 154, 197 
\bibitem{b35} Wenzel, W., 1984, IBVS, 2481
\bibitem{b36} Wood J.H., Crawford C.S., 1986, MNRAS, 222, 645
\bibitem{b37} Wood J.H., Horne K., Berriman G., Wade R., O'Donoghue D.,
		Warner B., 1986, MNRAS, 219, 629
\bibitem{b38} Wood J.H., Horne K., Berriman G., Wade R., 1989, ApJ, 341, 974
\bibitem{b39} Wood J.H., Horne K., 1990, MNRAS, 242, 609
\bibitem{b40} Wood J.H., Irwin M.J., Pringle J.E., 1985, MNRAS, 214, 475

\end{thebibliography}
\end{document}